## Title

A Pathologist-Annotated Dataset for Validating Artificial Intelligence: A Project Description and Pilot Study

## Authors

Sarah N Dudgeon (1), Si Wen (1), Matthew G Hanna (2), Rajarsi Gupta (3), Mohamed Amgad (4), Manasi Sheth (5), Hetal Marble (6), Richard Huang (6), Markus D Herrmann (7), Clifford H. Szu (8), Darick Tong (8), Bruce Werness (8), Evan Szu (8), Denis Larsimont (9), Anant Madabhushi (10), Evangelos Hytopoulos (11), Weijie Chen (1), Rajendra Singh (12), Steven N. Hart (13), Joel Saltz (3), Roberto Salgado (14), Brandon D Gallas (1)

**Affiliations of Authors:**

((1) United States Food and Drug Administration, Center for Devices and Radiologic Health, Office of Science and Engineering Laboratories, Division of Imaging Diagnostics & Software Reliability, White Oak, MD, (2) Memorial Sloan Kettering Cancer Center, New York, NY, (3) Stony Brook Medicine Dept of Biomedical Informatics, Stony Brook, NY, (4) Department of Pathology, Northwestern University, Rubloff Building, 750 N Lake Shore, Chicago Illinois 60611, (5) United States Food and Drug Administration, Center for Devices and Radiologic Health, Office of Product Quality and Evaluation, Office of Clinical Evidence and Analysis, Division of Biostatistics, White Oak, MD, (6) Massachusetts General Hospital/Harvard Medical School, Boston, MA, (7) Computational Pathology, Department of Pathology, Massachusetts General Hospital and Harvard Medical School, Boston, MA, (8) Arrive Origin, San Francisco, CA, (9) Department of Pathology, Institut Jules Bordet, Brussels, Belgium, (10) Case Western Reserve University, Cleveland, OH, (11) iRhythm Technologies Inc., San Francisco, CA, (12) Northwell health and Zucker School of Medicine, New York, NY, (13) Department of Health Sciences Research, Mayo Clinic, 200 1st St. SW, Rochester MN, (14) Division of Research, Peter Mac Callum Cancer Centre, Melbourne, Australia; Department of Pathology, GZA-ZNA Hospitals, Antwerp, Belgium)

**Corresponding Author:**

Brandon D. Gallas, PhD, Mathematician
Division of Imaging, Diagnostics, and Software Reliability (DIDSR)
Office of Science and Engineering Laboratories (OSEL)
Center for Devices and Radiological Health (CDRH)
US Food and Drug Administration
10903 New Hampshire Ave, WO-62 Rm 4104, Silver Spring, MD 20993
Email: brandon.gallas@fda.hhs.gov

## Article Keywords

AI Validation, Reference Standard, Medical Image Analysis, Pathology, Tumor Infiltrating Lymphocytes

## Abstract

Purpose:

Validating artificial intelligence (AI) algorithms for clinical use in medical images is a challenging endeavor due to a lack of standard reference data (ground truth). This topic typically occupies a small portion of the discussion in research papers since most of the efforts are focused on developing novel algorithms. In this work, we present a collaboration to create a validation dataset of pathologist annotations for algorithms that process whole slide images (WSIs). We focus on data collection and evaluation of algorithm performance in the context of estimating the density of stromal tumor infiltrating lymphocytes (sTILs) in breast cancer.

Methods:

We digitized 64 glass slides of hematoxylin- and eosin-stained ductal carcinoma core biopsies prepared at a single clinical site. A collaborating pathologist selected 10 regions of interest (ROIs) per slide for evaluation. We created training materials and workflows to crowdsource pathologist image annotations on two modes: an optical microscope and two digital platforms. The microscope platform allows the same ROIs to be evaluated in both modes. The workflows collect the ROI type, a decision on whether the ROI is appropriate for estimating the density of sTILs, and if appropriate, the sTIL density value for that ROI.

Results:

In total, 19 pathologists made 1,645 ROI evaluations during a data-collection event and the following two weeks. The pilot study yielded an abundant number of cases with nominal sTIL infiltration. Furthermore, we found that the sTIL densities are correlated within a case, and there is notable pathologist variability. Consequently, we outline plans to improve our ROI and case sampling methods. We also outline statistical methods to account for ROI correlations within a case and pathologist variability when validating an algorithm.

Conclusion:

We have built workflows for efficient data collection and tested them in a pilot study. As we prepare for pivotal studies, we will consider what it will take for the dataset to be fit for a regulatory purpose: study size, patient population, and pathologist training and qualifications. To this end, we will elicit feedback from the FDA via the Medical Device Development Tool program and from the broader digital pathology and AI community. Ultimately, we intend to share the dataset, statistical methods, and lessons learned.

## Introduction

Artificial intelligence is often used to describe machines or computers that mimic "cognitive" functions associated with the human mind, such as "learning" and "problem-solving" [1]. Machine learning (ML) is an artificial intelligence (AI) technique that can be used to design and train software algorithms to learn from and act on data. Although AI/ML has existed for some time, recent advances in algorithm architecture, software tools, hardware infrastructure, and regulatory frameworks have enabled healthcare stakeholders to harness AI/ML as a medical device. Such medical devices have the potential to offer enhanced patient care by streamlining operations, performing quality control, supporting diagnostics, and enabling novel discovery.

While AI/ML has already found utility in radiology, the role of AI/ML algorithms in pathology has been a matter of wide discussion [2]–[7]. Recent technological advancements and market access of systems that scan glass slides to create digital whole slide images (WSIs) have opened the door to a myriad of opportunities for AI/ML applications in digital pathology [8], [9]. While pathology is new to digitization, the field is expected to extend algorithms to a broad range of clinical decision support tasks. This technology shift is reminiscent of the digitization of mammography in 2000 [10] and the first computer-aided detection device (CADe) in radiology in 1998, the R2 ImageChecker [11]. The R2 CADe device marked regions of interest likely to contain microcalcifications or masses, initially evaluating digitized screen-film mammograms rather than digital acquisition of mammography images.

Fourteen years after the R2 ImageChecker was approved by the US FDA, regulatory guidance for CADe was finalized in two documents. While both guidance documents are specific to radiology, their principles are applicable to other specialties, including digital pathology. The first document generally delineates how to describe a CADe device and assess its "stand-alone" performance [12]. In the pathology space, this might be referred to as analytical validation. The second guidance document covers clinical performance assessment, or clinical validation [13]. The document was recently updated and discusses issues such as study design, study population, and the reference standard. Related issues are also discussed in a paper summarizing a meeting jointly hosted by the FDA and the Medical Imaging Perception Society [14].

Regardless of the technology providing the data or the algorithm architecture, software as a medical device (SaMD) must be analytically and clinically validated to ensure safety and effectiveness before clinical deployment [15]. One critical aspect of validation is to assess its accuracy: compare algorithm predictions to true labels using hold-out validation data, data that are independent from data used during development. Validation data includes patient data (features) on which the algorithm will make predictions as well as the corresponding reference standard (ground truth or label). The reference standard can be established by using an independent "gold standard" modality, longitudinal patient outcomes, or when these are not available or appropriate, a reference standard established by human experts. What constitutes the "ground truth" and how to approach it is a topic of discussion even in more traditional diagnostic test paradigms, and certainly so in evolving areas such as SaMD.

In this work we focus on the often-challenging task of establishing a reference standard using pathologists. The "interpretation by a reviewing clinician" is listed as a reference standard in the radiology CADe guidance documents and acts as the reference standard (in full or in part) in many precedent-setting radiology applications [16, p. 170], [17, p. 170], [18]. In pathology, the reference standard for evaluating performance in the Philips IntelliSite Pathology Solution (PIPS) regulatory submission, "was based on the original sign-out diagnosis rendered at the institution, using an optical (light) microscope" [8].

In this manuscript, we present a collaborative project to produce a validation dataset established by pathologist annotations. The project will additionally produce statistical analysis tools to evaluate algorithm performance. The context of this work is the validation of an algorithm that measures, or estimates, the density of tumor infiltrating lymphocytes (TILs), a prognostic biomarker in breast cancer. Given the cross-disciplinary nature of the study, the volunteer effort comprises an international, multidisciplinary team working in the pre-competitive space. Project participants include the FDA Center for Devices and Radiological Health's Office of Science and Engineering Laboratories (OSEL), clinician-scientists from international health systems, academics, professional societies, and medical device manufacturers. By incorporating diverse stakeholders, we aim to address multiple perspectives and emphasize interoperability across platforms.

We are pursuing qualification of the final validation dataset as an FDA Medical Device Development Tool (MDDT) [19]. In doing so, we have an opportunity to receive feedback from an FDA review team while building the dataset. If the dataset qualifies as an MDDT, it will be a high-value public resource that can be used in AI/ML algorithm submissions, and our work may guide others to develop their own validation datasets.

Definitions of terms in AI-based medical device development and regulation are evolving. For example, there have been inconsistent usage of "testing" vs "validation". To avoid this confusion, we are refer to building, training, tuning, and validating algorithms, where tuning is for hyperparameter optimization, and validation is for assessing or testing the performance of AI/ML algorithms. There is also some confusion between the terms "algorithm" and "model". In this work we will use the term "algorithm" to refer to the SaMD, the device, the software that is or will be deployed. Some may refer to the SaMD as the "model", but we shall use "model" to refer to the description of the algorithm (the architecture, image normalization, transfer learning, augmentation, loss function, training, hyperparameter selection, etc).

Herein, we present our efforts to source a pathologist-driven reference standard and apply it to algorithm validation, with an eye toward generating a fit-for-regulatory-purpose dataset. Specifically, we review the clinical association between TILs and patient outcomes in the context of accepted guidelines for estimating TIL density in tumor-associated stroma (sTIL density). We then imagine an algorithm that similarly estimates sTIL density and could use an sTIL density annotated dataset for validation. Next, we describe the breast cancer tissue samples used in our pilot study, the data-collection methods and platforms, and the pathologists we recruited and trained to provide sTIL density estimates in regions of interest (ROIs) using digital and microscope platforms. We also present some initial data and outline how we plan to account for pathologist variability when estimating algorithm performance.

## Technical Background
### Tumor Infiltrating Lymphocytes (TILs)

Tumor Infiltrating Lymphocytes (TILs) are an inexpensively-assessed, robust, prognostic biomarker that is a surrogate for anti-tumor, T cell-mediated immunity. Clinical validity of TILs as a prognostic biomarker in early-stage, triple negative breast cancer (TNBC), as well as in HER2+ breast cancer, has been well-established via Level 1b evidence [20]–[23]. Two pooled analyses of TILs, in the adjuvant setting for TNBC [21] and neoadjuvant setting across BC-subtypes [22], included studies that have evaluated TILs in archived tissue samples based on published guidelines [24]. Incorporating TILs into standard clinical practice for TNBC is endorsed by international clinical and pathology-standards (St. Gallen 2019 recommendation; WHO2019 recommendation; ESMO2019 recommendation) [25]–[28]. It is expected that TILs will be assessed to monitor treatment response in the future [29], [30]. Further, evidence is emerging that TIL-assessment will be done in other tumor types as well, including melanoma, gastrointestinal tract carcinoma, non-small cell lung carcinoma and mesothelioma, and endometrial and ovarian carcinoma [31], [32].

### Visual and Computational TIL Assessment

Given the recent and evolving evidence of the prognostic value of TIL assessment, there have been several efforts to create algorithms to estimate TIL density in cancer tissue. Amgad et al. provide an excellent summary of this space, including a table of algorithms from the literature, an outline with visual aids for TIL assessment, as well as a discussion on validation and training issues [32], [33]. While some algorithms are leveraging details about the spatial distribution of individual TILs in different tissue compartments [34]–[36], the guidelines for pathologists are to calculate the sTIL density [24] defined as the area of sTILs divided by the area of the corresponding tumor-associated stroma.

In this work we imagine an algorithm that estimates the density of sTILs in pathologist-marked ROIs in WSIs of hematoxylin- and eosin-stained slides (H&E) containing breast cancer needle core biopsies. Amgad et al. refer to these quantitative values as computational TIL assessments (CTA) and visual TIL assessments (VTA), respectively. Such an algorithm produces quantitative values [37] that are equivalent to those proposed in the guidelines for pathologists. This provides the opportunity for using pathologist evaluations as the reference standard for such an algorithm.

We propose the following clinical workflow: (1) Patient imaging finds an abnormality suspected for breast cancer. Physicians order a needle core biopsy to assess the tissue. (2) TILs will be scored during histopathologic evaluation and diagnosis. Specifically, pathologists will score the TILs in each H&E-stained breast cancer core-biopsy with assistance from an algorithm. Or, depending on the algorithm intended use, the sTIL score could be created automatically, without pathologist input. (3) The sTIL density will then be reported in the patient's pathology report.

### Algorithm Validation

Before it can be marketed and applied in the clinical workflow, any algorithm/software as a medical device should be well validated. Validation of algorithms for clinical use follows the building, training, and tuning phases of algorithm development. There are two main categories of algorithm validation: analytical and clinical. For both categories, a reference standard is needed. For algorithms that evaluate WSIs of H&E slides, there are generally three kinds of truth: patient outcomes, evaluation of the tissue with other diagnostic methods, and evaluation of the slide by pathologists. This work focuses on truth as determined by pathologists.

Analytical validation, or stand-alone performance assessment, focuses on the precision and accuracy of the algorithm, and compares algorithm outputs directly against the reference standard (Figure 1-A). In a clinical validation study, the algorithm end user, here a pathologist, evaluates cases without and with the algorithm outputs; Figure 1-B shows an independent-crossover clinical validation study design. There is typically a washout period between the evaluations by the same pathologists evaluating the same cases without and with the algorithm outputs, where the order in which these viewing modes are executed is randomized and balanced across pathologists and batches of cases. Figure 1-C shows a putative sequential clinical validation study design for an algorithm intended to be used as a decision-support tool after the clinician makes their conventional evaluation. We have depicted two populations of pathologists in our proposed clinical validation studies: experts for establishing the reference standard and end users for evaluating performance without and with the algorithm outputs.

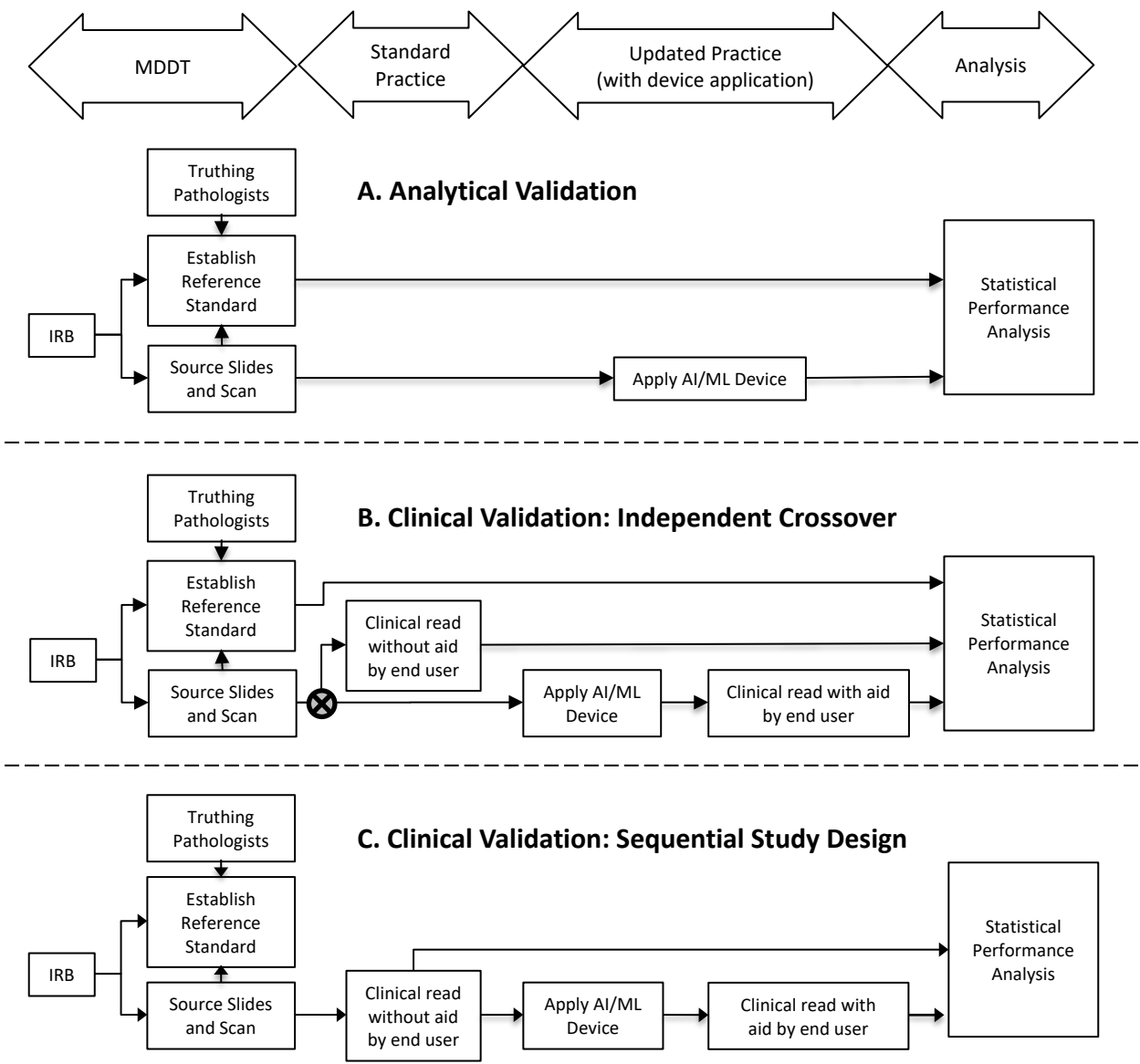

*Figure 1:* A. Study design for Analytical Validation of an algorithm (stand-alone performance assessment). Algorithm outputs are compared to the reference standard. B. Independent Crossover study design for Clinical Validation has two arms corresponding to pathologist evaluations without and with the algorithm. We compare the performance of these two evaluation modes. C. Sequential study design for Clinical Validation has one arm corresponding to end-user evaluations first without and then with the algorithm as an aid. A comparison is made between the performance of these two evaluation modes.

Current best practice for algorithm validation is to source slides from multiple independent sites different from the algorithm-development site to ensure algorithm generalizability, also known as external validation [38]–[41]. Developers should also be blinded to the validation data prior to a validation study, eliminating potential bias arising from developers training to the test [42]–[45]. These practices generally assume that the algorithm is locked; the architecture, parameters, weights, and thresholds should not be changed before the algorithm is released into the field. Validation of algorithms that are not locked – algorithms that rely on 'active learning' and 'online' learning, or hard negative mining, where the training is done iteratively and continuously – is an area that is still evolving and not in the scope of this work [32], [46]–[50].

## Approach
### Data – Pathology Tissue and Images

We, through a partnership with the Institut Jules Bordet, Brussels, Belgium sourced 77 matched core biopsies and surgical resections. Of these cases, 65 were classified as invasive ductal carcinoma and 12 were invasive lobular carcinoma. There was no patient information provided with these slides, no metadata such as age, race, cancer stage, or subtype (morphologic or molecular). This study was approved by the Ethics Commission of the Institut Jules Bordet.

The slides are 2019 re-cuts of formalin-fixed, paraffin-embedded tissue blocks from a single institution. Slide preparation was performed at the same institution by a single laboratory technician. Specifically, one 5 um-thick section was mounted on a glass slide and stained with hematoxylin and eosin (H&E). The slides were scanned on a Hamamatsu Nanozoomer 2.0-RS C10730 series at 40x equivalent magnification (scale: 0.23 um per pixel).

For our pilot study, we included eight batches of eight cases each; a case refers to the slide-image pair. The remaining 13 slides were not used for the pilot study. All 64 cases were biopsies of invasive ductal carcinomas; no resection specimens were used. Batches split data collection into manageable chunks for pathologists. Each batch was expected to take about 30 minutes to annotate. Batches also allowed us to make assignments for pathologists that help distribute evaluations across all cases and ROIs. We targeted 5 pathologist evaluations per ROI for the pilot study.

### Data Collection = ROI Annotation

Data collection, or ROI annotation, is broken into ROI selection and ROI evaluation in this work. ROI selection is a data curation step preceding ROI evaluation. The purpose of selecting ROIs ahead of ROI evaluation is to allow multiple pathologists to evaluate the same ROIs quickly. For our pilot study, ROI selection was performed by a collaborating pathologist using the digital platforms. Subsequent ROI evaluation was performed by recruited pathologists using digital and microscope platforms. The platforms, ROI selection and evaluation, and the pathologists that participated in the pilot study are described in more detail below.

## Digital Platforms

For this work, we have two digital platforms for viewing and annotating WSIs: PathPresenter [51] and caMicroscope [52]. Screen shots of the user interfaces are shown in Figure 2-A&B. Pathologists can log in from anywhere in the world, and annotate images using web-based viewers.

Leadership from PathPresenter and caMicroscope are collaborators in this project and have supported the development of controlled and standardized workflows to select ROIs and to evaluate ROIs. Both platforms can read and write annotations using the ImageScope XML format [53], and we have used that format to share ROIs and create an identical study on both platforms. Both platforms also record the pixel width and height and the zoom setting of the WSI area being viewed. We have not yet imposed display requirements in the pilot study but that will be discussed for future phases of our project.

Using more than one platform, including the microscope platform described next, allows us to involve more partners that can provide different perspectives, build redundancy to mitigate against a collaborator leaving the team, and promote interoperability as we progress to future phases of the project. The validation dataset will be based on the microscope platform, and the digital platforms allow fast development and understanding of our study and also allow us to compare microscope-mode to digital mode evaluations.

## Microscope Platform

The microscope platform we use is a hardware and software system called eeDAP, an Evaluation Environment for Digital and Analog Pathology [54]. The system uses a computer-controlled motorized stage and digital camera mounted to a microscope. eeDAP software registers the location of what is seen in the physical tissue through the microscope to the corresponding location in a WSI. Registration is accomplished through an interactive process that links the coordinates of the motorized stage to the coordinates of a WSI image. Registration enables the evaluation of the same ROIs in both the digital and microscope domains.

Similar to the digital platforms, the eeDAP software includes a utility to read and write ImageScope XML files, and a graphical user interface implementing the ROI evaluation workflow (See Figure 2-C) [55]. A research assistant supports the pathologist by entering data into the eeDAP GUI and monitoring registration accuracy. The square ROI is realized with a reticle in the eyepiece. As annotations are collected on the slide, they are scanner-agnostic and may be mapped to any scanned version of the slide using the eeDAP registration feature.

**A.**

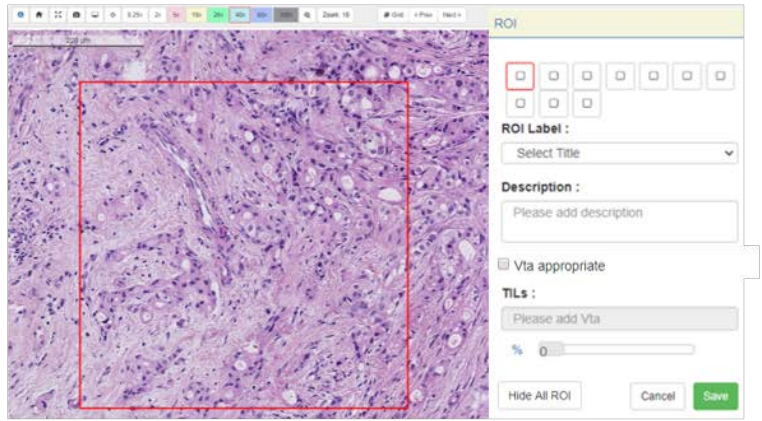

**B.**

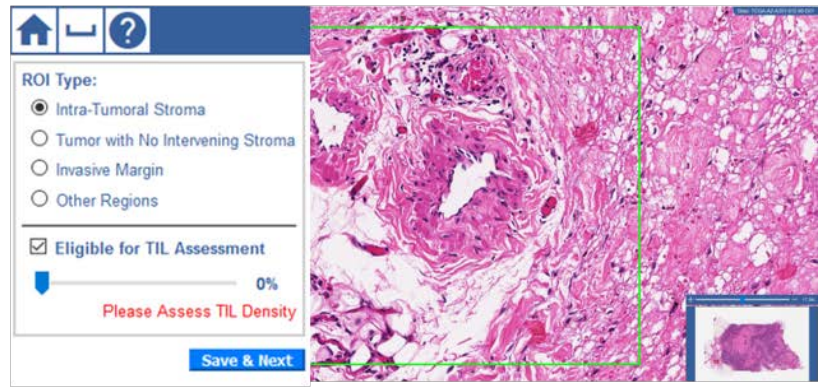

**C.**

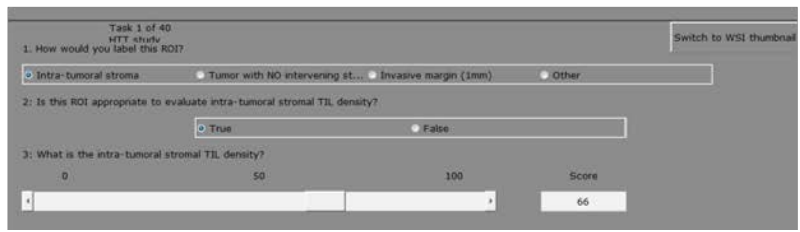

*Figure 2*: Screenshots from graphical user interfaces (GUI) of three platforms used in data collection. All three collect a descriptive label of the ROI (Table 1), a binary evaluation of whether the ROI is appropriate for sTIL density estimation, and an estimate of sTIL density via slider bar or keyboard entry. A. PathPresenter and B. caMicroscope are digital platforms. C. Evaluation Environment for Digital and Analog Pathology (eeDAP) microscope platform. In data collection, the pathologist is at the microscope, while a study coordinator records evaluations through the GUI.

## ROI Selection: Study Preparation

A board-certified collaborating pathologist marked 10 ROIs on each of the 64 cases using the digital platforms described above. The ROIs were 500 um x 500 um squares. The instructions were to target diverse morphology from various locations within the slide. More specific instructions were to target areas with and without tumor-associated stroma, areas where sTIL densities should and should not be evaluated. More details on selecting specific ROI types can be found in Table 1. An algorithm is expected to perform well in all these areas, so it is vital that the dataset include them.

## Table 1: ROI types (labels)

| |
|---|
| **<u>Intra-tumoral stroma (aka tumor-associated stroma): Select ~3 ROIs</u>** |
| ● Be sure to include regions with lymphocytes (TILs) |
| ● If there are lymphocytic aggregates, make sure to capture both lymphocyte-depleted and lymphocyte-rich areas within the same ROI if possible |
| ● Preferable to include some tumor in the same ROI — i.e. carcinoma cells as well as their associated stroma |
| ● If variable density within the slide, make sure to capture ROIs from different areas with different densities |
| **<u>Invasive margin (Tumor-stroma transition): Select ~2 ROIs</u>** |
| ● If heterogeneous tumor morphology, sample from different tumor-stroma transitions for each |
| **<u>Tumor with no intervening stroma: Select ~2 ROIs, if possible</u>** |
| ● If heterogeneous tumor morphology, sample from different morphologies |
| ● Be sure to sample from: vacuolated tumor cells, dying tumor cells, regions of different densities of tumor |
| ● Will be used to capture/assess intra-tumoral TILs and/or detect false positive TIL detections in purely cancerous regions. |
| **<u>Other regions: Select ~3-4 ROIs</u>** |
| ● ~1 from "empty"/distant/uneventful stroma |
| ● ~1 from hyalinized stroma, if any |
| ● ~2 other regions: |
|     ○ Necrosis transition (including comedo pattern) |
|     ○ Normal acini/ducts |
|     ○ Blood vessels |
|     ○ Others at pathologist discretion |

## ROI Evaluation

In current project protocols, we crowdsource pathologists to participate in ROI evaluation, separate from the pathologist who completed ROI selection. These pathologists will first label the ROI by one of the four labels given in Table 1. Pathologists then mark if the ROI is appropriate for evaluating sTIL density. This question is designed to determine if the area has tumor-associated stroma or not. If there is no tumor-associated stroma, annotation is complete. If there is tumor-associated stroma, the pathologist needs to estimate the density of TILs appearing in the tumor-associated stroma. The platforms allow integers 0 to 100, with no binning or thresholds. The motivation is to allow for thresholds to be determined later as the role of TILs becomes more clear and patient management guidelines are developed.

## Pathologist Participants in ROI Evaluation

Pathologist participants were recruited at a meeting of the Alliance for Digital Pathology immediately preceding the February 2020 USCAP annual meeting [6]. That meeting launched the in-person portion of pilot phase data collection. Board-certified anatomic pathologists and anatomic pathology residents were eligible to participate. To participate, they were asked to review the informed consent [56] and the training materials: the guidelines on sTIL evaluation [24] and a video tutorial and corresponding presentation about sTIL evaluation, the project, and using the platforms [57]. Reviewing the sTIL evaluation training was required before participating and took about 30 minutes. Pathologists were asked to label the ROI according to the types given in Table 1, a True-False decision about whether sTIL densities should or should not be evaluated, and if True, an estimate of the sTIL density.

In total, 19 pathologists made 1,645 ROI evaluations during the February event and the two weeks following. The primary platform at the event was the eeDAP microscope system where 7 pathologists made 440 evaluations. Most of the evaluations made on the digital platforms were made by pathologists who could not attend in person. Data collection in digital mode took approximately 30-40 minutes per batch and twice that long in microscope mode. The increased time for microscope evaluation was due to the motorized stage movements.

## Reference Standard (Truth) from Pathologists

The sTIL density measurements from pathologists are subject to bias and variance due to differences in pathologist expertise and training. In this work, we collected observations from multiple pathologists for each ROI, and then we averaged over the pathologists. While the precision of these values can be estimated, averaging over pathologists ultimately ignores pathologist variability in the subsequent algorithm performance metric. As such, we also let the observations from each pathologist stand as noisy realizations of the truth. This approach is used in related research on inferring truth from the crowd for the purpose of training an algorithm [58]. For our work, however, the purpose is to properly account for pathologist variability when estimating the uncertainty of algorithm performance.

### Performance Metric For sTIL Density Values

The primary endpoint of an algorithm that produces quantitative values needs to measure how close the values from the algorithm ($Predicted_i$) are to the reference standard ($Truth_i$). To evaluate "closeness", one appropriate performance metric that we are focusing on is the Root Mean Squared Error (RMSE):

$$\widehat{RMSE} = \sqrt{\frac{1}{N}\sum_i^N (Predicted_i - Truth_i)^2}\,, \tag{1}$$

where N is the number of ROIs. Smaller values of $\widehat{RMSE}$ indicate predicted values are closer to the truth, and thus better algorithm performance. Eq. 1 shows the RMSE estimated from a finite population (e.g. a finite sample of ROIs). As we consider a statistical analysis for our work – estimating uncertainty, confidence intervals, and hypothesis tests – we look to the infinite population quantity without the square root [59]–[62]:

$$MSE = E[(Predicted_i - Actual_i)^2] = bias^2 + variance. \tag{2}$$

Here we see that MSE measures accuracy and precision, similar to Lin's concordance correlation coefficient [63].

There are two main challenges to analyzing the differences between predictions and truth in our work. First, the sum in Eq. 1 is really a sum over ROIs nested within cases. These values are not independent and identically distributed (iid), as is generally assumed for Eq. 1. There should be a subscript for both case and ROI, and the statistical analysis needs to account for the correlation between values from ROIs within a case. In Figure 3 we show that sTIL densities are not iid across cases. The data are from one pathologist evaluating three cases that have different levels of sTIL infiltration. We see the sTIL densities are correlated within a case, and the variance is increasing with the mean. The distribution of sTIL densities is not the same for every case.

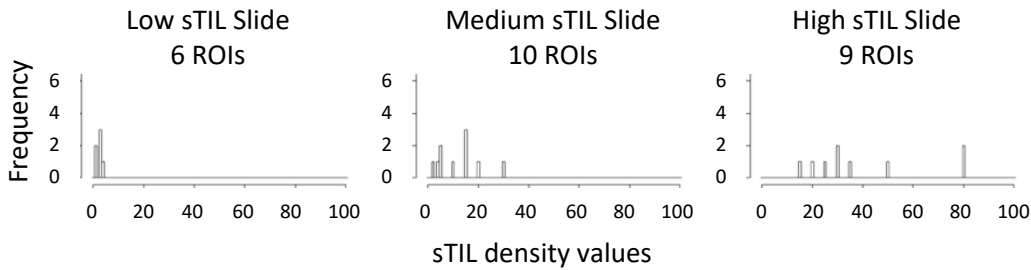

*Figure 3:* In this figure we show the distribution of sTIL densities in three slides with different levels of infiltration: A) Low, B) Medium, C) High. The sTIL densities were from one pathologist. As not all ROI labels are appropriate for sTIL density evaluation, not every case will contain TIL evaluations for all 10 ROIs.

The second challenge in our work is to account for the variability from pathologist to pathologist. This variability can be seen in Figure 4, which is a scatter plot showing the paired sTIL densities from two pathologists. Our strategy for addressing pathologist variability is to replace the single reference score in Eq. 1 with pathologist-specific values.

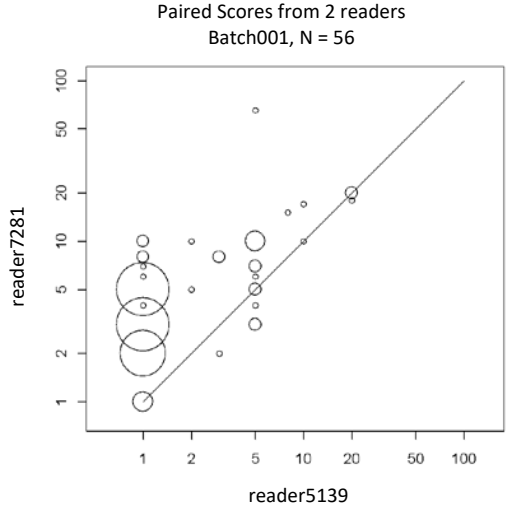

*Figure 4:* Scatter plot of sTIL densities from two pathologists on eight slides (one batch) that led to 56 paired observations. The plot is scaled by a log-base-10 transformation (with zero sTIL values changed to ones). The size of the circles is proportional to the number of observations at that point.

To address these two challenges, we rewrite Eq. 2 as

$$MSE = E\left[\left(Y_{kl} - X_{jkl}\right)^2\right], \tag{3}$$

where $X_{jkl}$ denotes the sTIL density from pathologist $j$ evaluating the $l^{th}$ ROI in case $k$ and $Y_{kl}$ denotes the sTIL density from the algorithm evaluating the $l^{th}$ ROI in case $k$. Furthermore, the expected value averages over pathologists, cases, and ROIs. It is this quantity that we wish to estimate, and we are developing such methods to account for the correlation of ROIs within a case and pathologist variability. The estimate may take the form of a summation over readers, cases, and ROIs, or it may be the result of a model that needs to be solved by more sophisticated methods that do not permit an explicit closed-form expression. The methods build on previous work on so-called multi-reader multi-case (MRMC) methods [64]–[67] and methods to evaluate intra- and inter-reader agreement [68].

## Discussion

The "High-Throughput Truthing" (HTT) moniker for this project reflects the data-collection methods as well as the spirit of the effort. The project was inspired by perception studies that have been run at annual meetings of the Radiological Society of North America [69]. Society meetings provide an opportunity to reach a high volume of pathologists away from the workload of their day job. A similar opportunity is available at organizations with many pathologists. We have explored both of these kinds of data collection opportunities via an event at the American Society of Clinical Pathology

Annual Meeting 2018 [70], [71] and an event at the Memorial Sloan Kettering Cancer Center [72], [73].

In addition to live events where we can use the eeDAP microscope system, our workflows on web-based platforms (PathPresenter and caMicroscope) can crowdsource pathologists from anywhere in the world. We have found these events to be low-cost, efficient opportunities to recruit pathologists and collect data. We plan to continue the project by scaling our efforts to a pivotal phase and disseminating our final validation dataset.

### FDA Medical Device Development Tool program

A key aim of this project is to pursue qualification of this dataset as a tool through the FDA Medical Device Development Tool program (MDDT) [19]. Pursuing qualification offers an opportunity to receive feedback from an FDA review team about building the dataset to be fit for a regulatory purpose. As we disseminate our work, we believe that this feedback will be valuable for the project and, more generally, for other public health stakeholders interested in the collection of validation datasets (industry, academia, health providers, patient advocates, professional societies, and government). A qualified tool has the potential to streamline the submission and review of validation data and allows the FDA to compare algorithms on the same pre-qualified data. In this way, the project may benefit the Agency and medical device manufacturers, as well as the larger scientific community.

The MDDT program was created by the FDA as a mechanism by which any public-health stakeholder may develop and submit a tool to the Agency for formal review. Tools are not medical devices. Rather, tools facilitate and increase predictability in medical device development and evaluation. Each tool is qualified for a specific **Context of Use** and may be used in a manufacturer's submission without needing to reconfirm its suitability and utility [19]. Qualified tools are expected to be made publicly available, which can include a licensing arrangement. In this way, qualified tools reduce burden to both the agency and the manufacturer and ultimately increase product quality and better patient outcomes. The proposed Context of Use for this work is given in Table 2.

## Table 2: Proposed Context of Use for an sTIL-density Annotated Dataset

The sTIL-density Annotated Dataset is a tool to be used to assess the accuracy of algorithms that quantify the density of stromal tumor infiltrating lymphocytes (sTILs). It is comprised of a dataset of slides, digital whole slide images, and annotations in regions of interest (ROIs). The annotations are compiled by pathologists using microscopes to evaluate glass slides of tissue samples from breast cancer needle core biopsies, where the tissue sections are stained with hematoxylin and eosin (H&E).

The exact platform and mechanisms for sharing the dataset have yet to be determined. However, the dataset will be shared broadly at no cost with any entity, subject to applicable terms required by either the FDA or the MDDT program. Possible terms would protect against data being used to "train to the test" using strategies such as data access via containers or data governance by written agreements. We can look to public challenges [74]–[77] to inform our data sharing plans and educational dissemination opportunities.

An MDDT dataset has the potential to significantly reduce the burden of manufacturers, especially small companies. Validation in the commercial space tends to be siloed, with each developer using distinct, licensed, and proprietary data. Our proposed MDDT may allow manufacturers and the FDA to avoid the time- and resource-consuming back-and-forth discussions to formulate a study design and protocol. Manufacturers may also be able to skip burdensome steps like obtaining Investigational Review Board (IRB) approvals, slide sourcing, reader recruitment, and collecting the data. Instead of planning statistical analyses from scratch, manufacturers may use the analyses developed from this project as an example to guide their work. These bypassed steps are represented in the column headings of Figure 1.

### Data Representativeness/Generalizability

A random set of breast-cancer biopsies are naturally expected to include the different immunophenotypic subtypes of TILs (CD4+, CD8+ T cells, natural killer cells) and a variety of shapes, locations, colors, and clustering of TILs [78]–[81]. Our current strategy of selecting ROIs gathers areas for sTIL evaluation with and without tumor-associated stroma, areas where sTIL densities should and should not be evaluated (Table 1). Despite efforts to assemble a balanced and stratified sample of ROI types, our pilot study data yielded an abundant number of cases with nominal sTIL infiltration. While this may be the true clinical distribution, for our MDDT we want to balance and stratify the sTIL density values across the expected range. For this, we intend to realize some data curation before ROI evaluation.

The MDDT dataset should also adequately represent the variability arising from pre-analytic differences (slide preparation) and the intended population (clinical subgroups). To sample pre-analytic differences, we intend to sample cases from at least three sites for the MDDT dataset. If possible, we will also create some cases that systematically explore the H&E staining protocol (incubation time, washing time, and stain strength).

There are several clinical subgroups that are appropriate to sample, like patient age, breast-cancer subtypes and stages [28], [82]–[84], and treatment at various time intervals. Sampling from all possible subgroups is challenging if not impossible. While our inclusion and exclusion criteria limit the use of our MDDT to a selective population, we do not expect to sample all the subgroups that might be required in an algorithm submission, and we do not expect to have the same metadata for all cases. It is important to note that while TILs are known to have the most prognostic value in certain molecular (genomic) subtypes (e.g. triple-negative, Her2 positive etc.), a TILs algorithm is most likely to be confounded by histologic subtype and characteristics. While there is some correspondence between genomic and histologic classifications of breast tumors, the histological presentation (morphology) of, say, a ductal carcinoma does not necessarily correlate well with its genomic composition. Any data that is not part of the MDDT but is required for a regulatory submission of an algorithm will ultimately be the responsibility of the algorithm manufacturer. We do not intend to sample treatment methods or longitudinal data.

### Pathologists and Pathologist Variability

In this work, our initial data shows notable variability in independent sTIL density estimates from multiple pathologists on each ROI (Figure 4), which is consistent with previous work in this area [32]. These findings further reinforce the need to collect data from multiple pathologists and the need to better understand this variability. We intend to explore the difference between averaging over pathologists and keeping them distinct when evaluating algorithm performance. In either case, we believe that a statistical analysis method should account for reader variability in addition to case variability. A final statistical analysis plan, including sizing the number of pathologists and cases, will be developed based on the pilot data, simulation studies, and feedback from the FDA's MDDT review team.

As we are crowdsourcing pathologists, we have received questions regarding the expertise of the participating pathologists. Initially, we accepted any board-certified pathologist or anatomic pathology resident, but the reader variability observed in the pilot data has caused us to reconsider. Improving the expertise of the pathologists would reduce pathologist variability and allow us to reduce the number of pathologists. Therefore, we are considering expanding our current training materials to include testing with an immediate feedback loop providing the reference standard for each ROI. We are also considering creating a proficiency test. The proficiency test could be built from the pilot study dataset. A robust training program could additionally serve the community beyond our specific project need.

As relates to the RMSE performance metric, which summarizes the bias as well as the variance of an algorithm, it is not clear whether the bias comes from the algorithm or the pathologist. Amgad et al. [85] found their algorithm to be biased low compared to the pathologists. They also found that the Spearman rank-based correlation was stronger for the algorithm-to-pathologist-consensus comparison compared to the pathologist-to-pathologist comparison (R=0.73 vs. R=0.66). The authors believe these results are related to pathologist bias and variability, and not the algorithm. While this may be true, it is difficult to know as only two pathologists provided sTIL density values. Furthermore, the comparison does not account for pathologist variability in either correlation result and is not an apples-to-apples comparison due to the consensus process. Still, it may be appropriate to consider other performance metrics, like Spearman's rank correlation or Kendall's tau, that treat pathologist sTIL density estimates as ordinal data rather than quantitative and calibrated data [68], [86]–[88].

### Relaunch and Future

While the live event portion of pilot phase data-collection was a burdensome process, we totaled 1,645 evaluations in ten hours. The live event was set up with four evaluation stations: 2 digital platforms and 2 microscope platforms. We created training materials and hosted an online training seminar before the event. We assembled recruitment materials and sent invitations to pathologists. We trained study administrators to operate eeDAP and assist pathologists with data collection at the microscope. All equipment was shipped and assembled on site. Data aggregation was completed via APIs. Not surprisingly, the data is stored quite differently on the two digital platforms, so we created scripts to clean and harmonize the raw data into common data frames. We began building a software package to analyze the clean data. In sum, the process took a lot of time and effort, but offered experiences to inform the next phase of our project.

To help pathologists improve their sTIL density estimates and collect more detailed data, we thought about what an algorithm generally would do: identify and segment the tumor, tumor-associated stroma, and sTILs. We thought it would be worthwhile to parallel these steps. We were already asking pathologists to label ROIs by tumor, margin, and the presence of tumor-associated stroma. We decided to also ask the pathologist to estimate the percent of the ROI area that contains tumor-associated stroma.

Data collection on the microscope system was put on hold because of the COVID-19 pandemic, but we relaunched data collection on the digital platforms in September 2020 to fill out observations across all batches of the pilot study. We invite Board-Certified pathologists to spend approximately 30 minutes on training and 30 minutes per batch on data collection [89]. With newly established agreements for sharing materials, we are in the process of securing more slides and images to sample the patient subgroups mentioned. We welcome parties that are able and willing to share such materials to contact us through the corresponding author. Similarly, we are looking for opportunities to set up high-throughput truthing events or find collaborating sites interested in hosting data collection events on their own. There are opportunities to set up their own eeDAP microscope system or borrow an existing system from us. We are willing to supervise and assist remotely.

## Conclusion

On the volunteer efforts of many and a nominal budget, we have created a team and a protocol, administrative materials and infrastructure for our high-throughput truthing project. We have sourced breast-cancer slides and crowdsourced pathologists. Our goal is to create an sTIL-density Annotated Dataset that is fit for a regulatory purpose. We hope this project can be a roadmap and inspiration for other stakeholders (industry, academia, health providers, patient advocates, professional societies, and government) to work together in the pre-competitive space to create similar high-value, fit-for-purpose, broadly accessible datasets to support the field in bringing algorithms to market and to monitor algorithms on the market.

## Acknowledgements

The HTT team acknowledges the work of collaborating pathology networks like the International Immuno-Oncology Biomarkers Working Group and the Alliance for Digital Pathology in their participant pathologist recruitment (www.tilsinbreastcancer.org & https://digitalpathologyalliance.org). R.S. is supported by a grant from the Breast Cancer Research Foundation (BCRF, grant No. 17-194) and J.S. is supported by grants from the NIH (UH3CA225021, U24CA180924). Ioanna Laios and Ligia Craciun contributed technical support and expertise in the preparation of glass slides. Lastly, the team thanks their developers; Krushnavadan Acharya (PathPresenter), Nan Li (caMicroscope), and Qi Gong (eeDAP).